\documentclass[10pt,a4paper]{article}
\usepackage[T1]{fontenc}       
\usepackage{amssymb}
\usepackage{amsmath}
\usepackage{braket}
\usepackage{graphicx,caption}
\usepackage{multicol}
\usepackage{scrextend}
\usepackage{lineno}
\usepackage{lettrine}
\usepackage{pdfpages}
\usepackage[normalem]{ulem}
\renewcommand{\thesection}{\Roman{section}} 
\renewcommand{\thesubsection}{\thesection.\Alph{subsection}}

\usepackage{titlesec}
\titleformat{\section}
  {\normalfont\fontsize{10}{15}\bfseries}{\thesection}{1em}{}
\titleformat{\subsection}
  {\normalfont\fontsize{10}{15}\bfseries}{\thesubsection}{1em}{}

\usepackage[utf8]{inputenc}
\usepackage{amsfonts}

\newcommand{\RR}[1]{\color{black}#1 \color{black}}

\graphicspath{{./Figures/}}

\usepackage[sort&compress,square]{natbib}
\setcitestyle{numbers}
\makeatletter
\newcommand{\overleftrightsmallarrow}{\mathpalette{\overarrowsmall@\leftrightarrowfill@}}
\newcommand{\overrightsmallarrow}{\mathpalette{\overarrowsmall@\rightarrowfill@}}
\newcommand{\overleftsmallarrow}{\mathpalette{\overarrowsmall@\leftarrowfill@}}
\newcommand{\overarrowsmall@}[3]{%
	\vbox{%
		\ialign{%
			##\crcr
			#1{\smaller@style{#2}}\crcr
			\noalign{\nointerlineskip}%
			$\m@th\hfil#2#3\hfil$\crcr
		}%
	}%
}
\def\smaller@style#1{%
	\ifx#1\displaystyle\scriptstyle\else
	\ifx#1\textstyle\scriptstyle\else
	\scriptscriptstyle
	\fi
	\fi
}
\makeatother
\newcommand{\tensor}[1]{\overleftrightsmallarrow{#1}}
\begin{document}
	\begin{center}
		
		{\large{\bf
			Electrical control of the $g$-tensor of the first hole 
			\\in a silicon MOS quantum dot\\
			}}
			
			
			\vskip0.5\baselineskip
			
			{\bf
				S. D. Liles$^{1,\dagger}$, F. Martins$^{1,2}$, D. S. Miserev$^{1,3}$, A. A. Kiselev$^4$, I. D. Thorvaldson$^{1}$, 
				M.~J.~Rendell$^{1}$, I. K. Jin$^{1}$, F. E. Hudson$^{5}$, M. Veldhorst$^{5,6}$, K. M. Itoh$^{7}$, O. P. Sushkov$^{1}$, 
				T.~D.~Ladd$^{1,4}$, A. S. Dzurak$^{5}$, A. R. Hamilton$^{1}$
			}
			
			\vskip0.5\baselineskip
			
			{\it
				$^{1}$School of Physics, University of New South Wales, Sydney NSW 2052, Australia\\
				$^{2}$Hitachi Cambridge Laboratory, J.J. Thompson Avenue, Cambridge CB3 0HE, United Kingdom\\
				$^{3}$Department of Physics, University of Basel, Klingelbergstrasse 82, CH-4056 Basel, Switzerland\\
				$^{4}$HRL Laboratories, LLC, 3011 Malibu Canyon Rd., Malibu, CA 90265, USA\\
				$^{5}$School of Electrical Engineering and Telecommunications, \\
				The University of New South Wales, Sydney NSW 2052, Australia\\ 
				$^{6}$QuTech and Kavli Institute of Nanoscience, TU Delft, 2600 GA Delft, The Netherlands\\
				$^{7}$School of Fundamental Science and Technology, Keio University, Yokohama, Japan\\
			}
			
			%
			
			\let\thefootnote\relax\footnote{$\dagger$ corresponding author - s.liles@unsw.edu.au\\}
\\
\textbf{Abstract}\vspace{1ex}\\ \parbox{0.8\textwidth}{
	   Single holes confined in semiconductor quantum dots are a promising platform for spin qubit technology, due to the electrical tunability of the $g$-factor of holes. However, the underlying mechanisms that enable electric spin control remain unclear due to the complexity of hole spin states. Here, we study the underlying hole spin physics of the first hole in a silicon planar MOS quantum dot. We show that non-uniform electrode-induced strain produces nanometre-scale variations in the HH-LH splitting. Importantly, we find that this \RR{non-uniform strain causes} the HH-LH splitting to vary by up to 50\% across the active region of the quantum dot. We show that local electric fields can be used to displace the hole relative to the non-uniform strain profile, allowing a new mechanism for electric modulation of the hole g-tensor. Using this mechanism we demonstrate tuning of the hole $g$-factor by up to 500\%. In addition, we observe a \RR{potential} sweet spot where d$g_{(1\overline{1}0)}$/d$V$ = 0, offering a configuration to suppress spin decoherence caused by electrical noise. These results open a path towards a previously unexplored technology: engineering of \RR{non-uniform} strains to optimise spin-based devices.} 
\end{center}	
				
\begin{multicols}{2}
\section{INTRODUCTION}
\lettrine{S}{ingle hole spins} confined in group IV quantum dots provide a promising path towards scalable quantum computing\cite{loss1998quantum,zwanenburg2013silicon,veldhorst2017silicon,scappucci2020germanium}. These devices can leverage well-established industrial platforms\cite{hutin2018si, pillarisetty2018qubit}, while also enabling rapid all-electric spin control\cite{golovach2006electric,bulaev2007electric,szumniak2012spin,maurand2016cmos}. 
Recent demonstrations have included single-qubit gate operations of holes in silicon devices\cite{maurand2016cmos} and up to four-qubit gate operations of holes Ge devices\cite{watzinger2018germanium, jirovec2021singlet,hendrickx2020fast,hendrickx2020single,hendrickx2021four}.

When developing spin qubit technology, a fundamental question arises: What defines the coupling of a single isolated spin to the external magnetic field? While this has been well studied for electrons \cite{zwanenburg2013silicon,hanson2007spins}, the complexity of hole spin states makes this a non-trivial question\cite{winkler2003spin,bulaev2005spin,chesi2014controlling,wang2016anisotropic,miserev2017mechanisms,miserev2017dimensional}. 
Holes occupy the valence band, which originates from $l$=1 atomic p-orbitals, with an effective spin of $S=\frac{3}{2}$ and a strong intrinsic spin-orbit coupling. For qubit devices, the combination of spin-orbit coupling and quantum confinement strongly modifies the hole spin properties, which become sensitive to the size and shape of the quantum dot\cite{schroer2011field,takahashi2013electrically}. 
In addition, the degree of mixing between the \RR{Heavy Hole} (\RR{HH},  $m_j$=$\pm$3/2) and \RR{Light Hole} (\RR{LH}, $m_j$=$\pm$1/2)  sub-bands leads to a mixed spin character. This causes holes spins to be very sensitive to effects \RR{that alter the HH-LH splitting,} such as crystal anisotropies, strain, and the confinement profile \cite{Chen2010,ares2013nature,miserev2017mechanisms}.

The $g$-tensor is the key parameter for studying the coupling of a spin-orbit state to a magnetic field\cite{zwanenburg2009spin,ares2013nature,van2014probing,srinivasan2016electrical,voisin2016electrical,bogan2017consequences,studenikin2019electrically,tanttu2019controlling}. 
However, most studies of the $g$-tensor of hole quantum dots have been performed using devices that confine an unknown number of holes\cite{crippa2018electrical,hu2012hole,ares2013sige,brauns2016electric,watzinger2016heavy,crippa2019gate,marx2020spin,geyer2021silicon}. 
This has hindered the ability to understand hole spin-qubit devices, since the number of holes is a primary factor influencing the orbital physics of the quantum dot\cite{kouwenhoven2001few}. It is imperative to know the quantum dot orbital wavefunction shape in order to make any quantitative comparison between experiments and theory, or to compare between different device designs or material systems.


\begin{figure*}[!tb]
	\centering
	\includegraphics[width=1.9\columnwidth ]{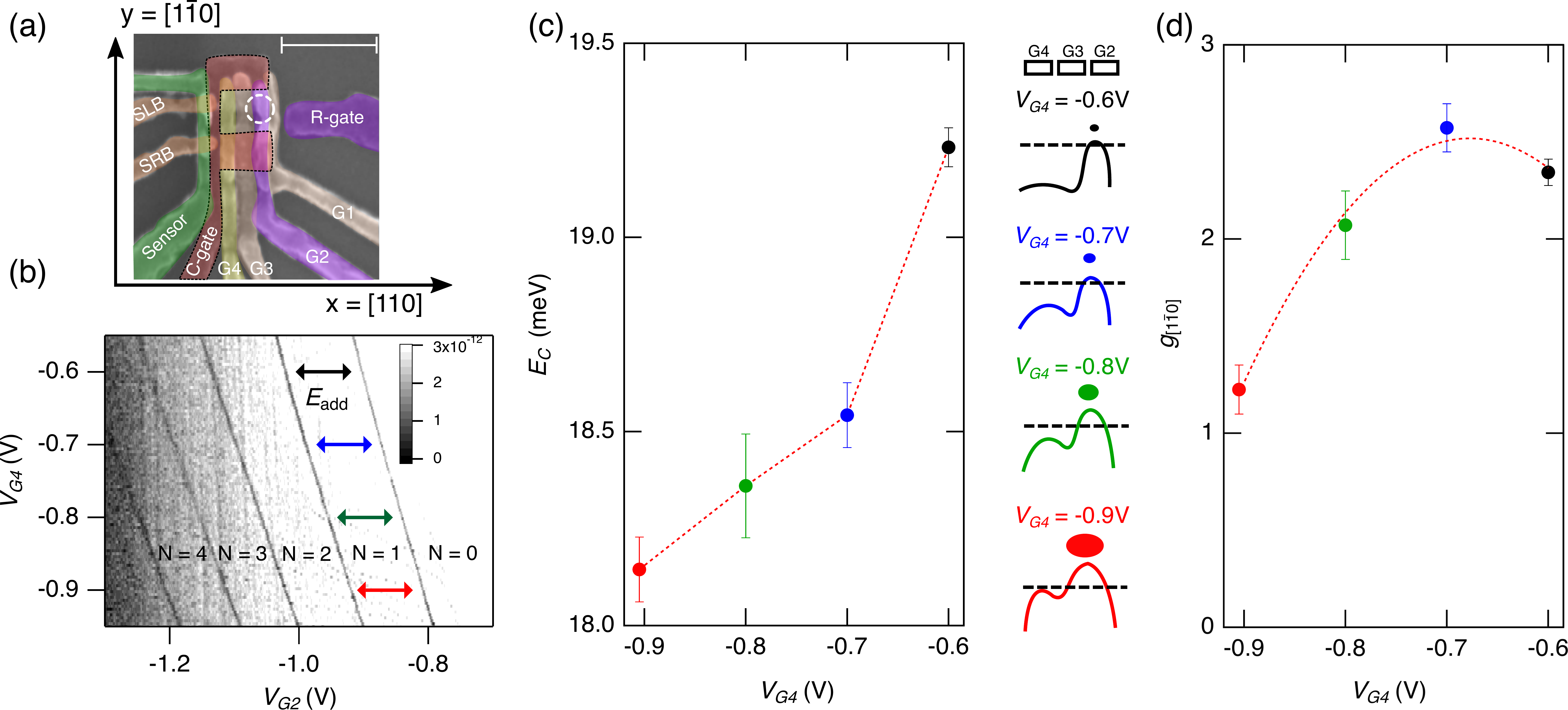}
	\caption[]{\small A single hole in a silicon quantum dot. (a) False colour SEM image of a device gate stack. A single quantum dot is formed in the region indicated by the white dashed circle by using gate G2 as the quantum dot plunger gate, while gates G1, G3 and the C-gate provide the electrostatic confinement. The in-plane crystal orientations are indicated, where the sample $x$-axis is the crystal axis [110], and the sample $y$-axis is the crystal axis [1$\overline{1}$0]. The out-of-plane direction is the sample $z$-axis, corresponding to crystal axis [001]. The horizontal white scale bar is 250~nm. (b) Charge stability diagram showing operation down to the last hole, where the gray-scale is d$I_{\text{sens}}$/d$V_{G2}$. A series of charge transitions can be observed as negative spikes in d$I_{\text{sens}}$/d$V_{G2}$. (c) The charging energy, $E_C$, measured at the $V_{G4}$ indicated by the colored horizontal arrows in (b). Schematics on the right indicate a line cut of the electrostatic potential energy along the sample $x$-axis. A hole quantum dot is formed at the potential maxima. The black horizontal dashed line indicates the Fermi energy, and the ellipse below each schematic represents single hole probability density, which is displaced and elongated as $V_{G4}$ finely tunes the local electrostatic environment. (d) The measured effective $g$-factor for a magnetic field applied along the sample $y$-axis [1$\overline{1}$0]. The dashed red line is a guide to the eye. 
	}
	\label{fig:StabilityDiagramFigure}
\end{figure*} 

Here we study the spin properties of the first hole confined in a planar silicon MOS quantum dot\cite{li2013single,spruijtenburg2013single,li2015pauli,liles2018spin}. 
By operating the device in the single charge ($N=1$) regime, we characterise the $g$-tensor in a known \RR{charge state with a well defined quantum dot orbital index.} This allows direct comparisons between experimental results and theoretical modeling, while the simple planar geometry allows the contributions of competing effects such as orbital alignment and \RR{non-uniform} electrode-induced strain to be separated \cite{thorbeck2015formation,park2016electrode}. 

Our results show that electrode-induced strain is key in mediating electric $g$-factor control in these hole MOS quantum dots. 
The effect of \RR{non-uniform} electrode-induced strain \RR{on the hole g-tensor} has not been previously considered for hole spin qubits. Therefore, these results open a new platform for hole spin-qubit technology, where the hole-spin qubits can be electrically manipulated by displacing the wave-function relative to precisely engineered \RR{electrode-induced} strain gradients.
\section{RESULTS}
\subsection{Isolating a single hole}
The device studied in this work was fabricated on an isotopically enriched $^{28}$Si  substrate with a high-quality 5.9~nm SiO$_2$ gate oxide. 
The device consists of a planar multi-layer aluminum gate stack\cite{angus2007gate,li2013single}. Figure \ref{fig:StabilityDiagramFigure}a shows a SEM image of the device layout. 
This layout allows the formation of a stable single hole quantum dot in the region indicated by the white circle\cite{liles2018spin}. 
The top gate of the adjacent charge sensor is indicated in green in Figure~\ref{fig:StabilityDiagramFigure}a. 
By independently monitoring the current through the charge sensor ($I_{\text{sens}}$), we can unambiguously identify the absolute number of holes occupying the quantum dot. 
Further details are provided in the methods section.

Figure~\ref{fig:StabilityDiagramFigure}b presents the charge stability diagram of the device, which was obtained by monitoring the transconductance (d$I_{\text{sens}}$/d$V_{G2}$) of the charge sensor.
The stability diagram shows a series of charge transitions, consistent with a single quantum dot formed under G2, with the number of holes indicated as $N$. 
Beyond the region labeled $N=0$, no further charge transitions were observed, confirming the device was operating down to the last hole. 

Figure~\ref{fig:StabilityDiagramFigure}c shows the Coulomb charging energy for the $N=1$ to $N=2$ transition, measured at different values of $V_{G4}$  (see the methods section for full details). 
For a fixed number of holes, the Coulomb charging energy ($E_{C}$) is inversely proportional to the size of the quantum dot confinement\cite{kouwenhoven2001few}. 
In Figure~\ref{fig:StabilityDiagramFigure}c, $E_{C}$ increases as $V_{G4}$ is made more positive, consistent with the quantum dot becoming smaller. 
Therefore, in the region between the first and second Coulomb peaks (indicated by the colored arrows in Figure~\ref{fig:StabilityDiagramFigure}b), it is possible to confine a single hole and use $V_{G4}$ to finely tune the spatial extent of the hole wavefunction. 

The remainder of this paper focusses on the $g$-factor of the first ($N=1$) hole confined in this planar quantum dot. 
The hole occupies the lowest energy orbital state, avoiding complications from higher quantum dot orbitals.

\subsection{Electrical modulation of the $N=1$ hole $g$-factor}
We examined the $g$-factor of the first hole as the shape of the wavefunction was systematically varied using the bias of a nearby electrode (G4). 
The effective hole $g$-factor was extracted from the linear increase in the $N=1$ addition energy with magnetic field $B$ (see S5 of Ref. \citenum{supp}). 
The in-plane magnetic field was aligned with the sample $y$-axis (crystal axis [1$\overline{1}$0]).  
Figure~\ref{fig:StabilityDiagramFigure}d presents the single hole effective $g$-factor ($g_{1\overline{1}0}$) for different electrostatic confinement profiles.
The magnitude of $g_{1\overline{1}0}$ can be tuned between 1.2$\pm$0.1 and 2.6$\pm$0.1 with only a small change of $V_{G4}$. 
From the maximum slope of Figure \ref{fig:StabilityDiagramFigure}d we obtain the maximum electric control over the $g$-factor as d$g$/d$V_{G4}$=8.1$\pm0.2$V$^{-1}$ (for this specific in-plane magnetic field orientation).
The observed $dg/dV_G$ for holes is six orders of magnitude larger than $dg/dV_G$ for electrons in identical silicon MOS devices\cite{tanttu2019controlling}, and is comparable to $dg/dV_G$ observed for holes in other Group-IV quantum dots\cite{crippa2018electrical,lawrie2020spin}.  
Based on the maximum $dg/dV_{G4}$ we estimate a \RR{minimum} Rabi frequency of 40~MHz\cite{kato2003gigahertz,ares2013sige,crippa2018electrical} \RR{(see methods section), however a full characterisation of the Rabi Frequency requires a more detailed study.\cite{crippa2018electrical}}

A key result of Figure~\ref{fig:StabilityDiagramFigure}d is the observation of a \RR{potential} ``sweet spot'' around $V_{G4}=-0.7$~V, where $dg_y/dV_{G4}=0$ 
\RR{(identification of a ``global sweet spot'' would require characterisation of dg/dV over the full gate parameter space).}
Sweet spots where $d\tensor{g}/dV_G=0$~V$^{-1}$ are important for qubits since the coupling between the spin and electric-fluctuations in the voltage source ($V_G$) are suppressed. 
Minimising the effects of charge noise is critical for hole spin qubits since the coherence time $T^*_2$ of hole spins in Group-IV quantum dots is primarily limited by electrical noise\cite{hendrickx2020single,kobayashi2020engineering}. Furthermore recent theoretical work shows that it is possible to engineer sweet spots where the dominant charge dephasing mechanism is suppressed while still allowing high speed electrical qubit control. \cite{wang2019suppressing,bosco2021hole,adelsberger2021hole}.

\begin{figure*}[!htbp]
	\centering
	\includegraphics[width=1.8\columnwidth]{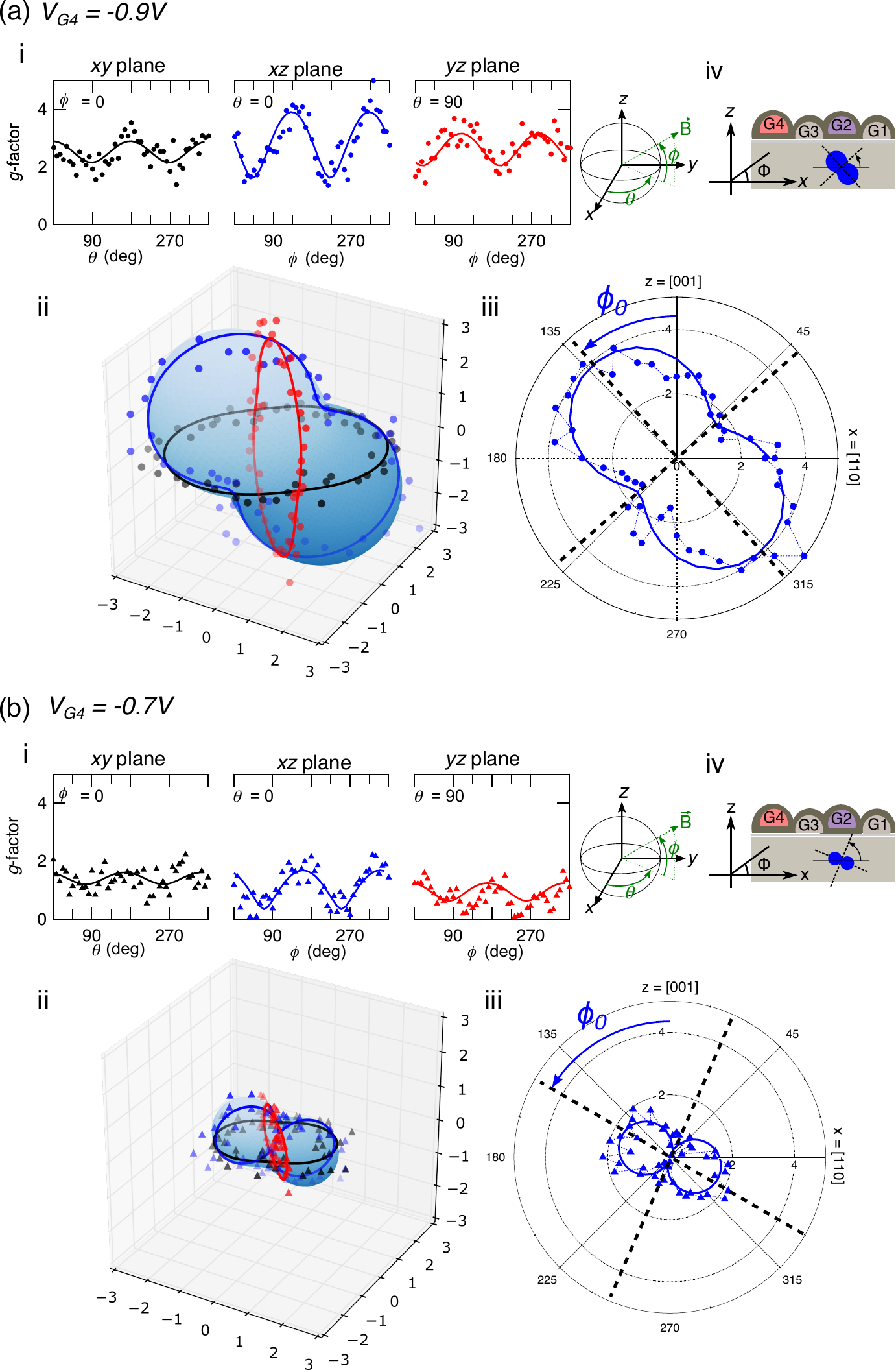}
	\caption[]{\small Electrical control of the $g$-tensor orientation. 
	Measurements of the hole $g$-tensor for (a) $V_{G4} = -0.9$~V and (b)  $V_{G4} = -0.7$~V. 
	(i) The measured effective $g$-factor for magnetic field rotations around the sample \RR{$z$} (black), \RR{$y$} (blue) and \RR{$x$} (red) axes. 
	Solid lines are a best fit of all data to Equation (\ref{eqn:gExp}). 
	\RR{Typical uncertainty in g-factor is 0.4, see S5.2. of Ref. \citenum{supp}}
	Angles $\theta$ and $\varphi$ define the orientation of the magnetic field $\vec{B}$ as indicated on the adjacent sphere.
	(ii) The shaded blue surface is the 3D $g$-tensor surface defined by the appropriate parameters in Table \ref{table:FittingParameters}. For reference the experimental data (circles/triangles) and best fit (solid lines) from (i) are included.
	(iii) Reproduces the $g$-factor measurements for the \RR{$y$-axis rotation (blue)} ($\theta = 0$) as a polar plot. 
	The radial axis is |$g$| and while the angle corresponds to $\varphi$.
	The axes of symmetry of the $g$-tensor in the $xz$ plane are indicated by the dashed black lines. 
	The magnitude of the tilting into the $x$ plane is indicated by the blue arrow, where the tilt in the $xz$ plane corresponds to $\phi_0$ of Equation (\ref{eqn:gExp}). 
	(iv) Shows the sample schematic with the $xz$ $g$-tensor surface, highlighting the titled $g$-tensor orientation with respect to the Si/SiO$_2$ interface.
	Triangles are used for raw data in \textbf{b}.
	}
	\label{fig:Figure2}
\end{figure*} 

\subsection{Characterising the $g$-tensor}
In hole systems, the coupling to a magnetic field is influenced by many factors, such as the 3D wavefunction shape, local strain, and the crystal anisotropy\cite{schroer2011field,takahashi2013electrically}. Fully defining the magnetic response to a magnetic field ($\vec{B}$) requires a $3\times3$ $g$-tensor with 6 free parameters. The $g$-tensor is defined as
\begin{equation}
\tensor{g} = R(\vartheta_0, \phi_0, \theta_0)
\begin{pmatrix}
g_{1}& 0 & 0 \\
0 & g_{2} & 0 \\
0 & 0 & g_{3}
\end{pmatrix} 
R^{-1}(\vartheta_0, \phi_0, \theta_0)
\label{eqn:gExp}
\end{equation} 
where the angles $\vartheta_0$, $\phi_0$, and $\theta_0$ define the orientation of the principal magnetic axes with respect to the sample $(x,y,z)$ axes, and $g_{1}$, $g_{2}$ and $g_{3}$ define the principal $g$-factors. 
The $R$ matrix represents the effect of three consecutive rotations around the sample axes (as described in the methods).

Experimental characterisation of the $g$-tensor requires measurements of the $g$-factor for a range of magnetic field orientations in all three dimensions. In this work, the $g$-tensor of the $N=1$ hole was characterised experimentally using a vector magnet. 
The magnetic field was fixed at $|\vec{B}| = 1$~T, and a \RR{$2\pi$ rotation in increments of  $\pi/24$ around the sample $x$, $y$ and $z$ axes was performed.} 
At each magnetic field orientation we extracted the $g$-factor from the linear change of the $N=1$ addition energy.
The measurement was repeated for two different confinement profiles, which were controlled by setting $V_{G4}=-0.9$~V (Figure~\ref{fig:Figure2}a-i) or $V_{G4}=-0.7$~V (Figure~\ref{fig:Figure2}b-i). 
The solid lines in Figure~\ref{fig:Figure2}a-i and Figure~\ref{fig:Figure2}b-i show the best fit of the full data set (all 144 points) to Equation~(\ref{eqn:gExp}).
The best fit parameters for both confinement profiles are presented in Table~\ref{table:FittingParameters}. 
Figures \ref{fig:Figure2}a-ii and \ref{fig:Figure2}b-ii present a 3D visualisation of the best fit $g$-tensor.  

For the case of $V_{G4}=-0.9$~V, the largest principal $g$-factor ($g_3$) is 3.9$\pm$0.1, and the smallest principal $g$-factor ($g_1$) is 1.4$\pm$0.2. 
The orientation of the $g$-tensor is distinctly tilted with respect to the sample axes (Figure \ref{fig:Figure2}a-ii), such that the principal magnetic axes are not aligned with any lithographic or crystallographic axes of the sample. 
To demonstrate this tilted orientation, Figure~\ref{fig:Figure2}a-iii shows the measured $g$-factor around the sample \RR{$y$} axis on polar axes. 
In particular we note that the $g$-tensor is tilted by $\phi_0=42^{\circ}\pm2^{\circ}$ in the $xz$ plane. 
Figure \ref{fig:Figure2}a-iv shows a schematic of the  $g$-tensor surface in the $xz$ plane of the sample, highlighting that the largest $g$-factor occurs when the magnetic field is tilted by $42^\circ$ away from the Si/SiO$_2$ interface. 
The observation that the principal axes of the $g$-tensor are not fixed by any sample axes is the key result of the $g$-tensor characterisation in Figure \ref{fig:Figure2}a.

We next investigate if the orientation of the $g$-tensor principal axes can be electrically tuned. 
The hole wavefunction shape was changed by varying $V_{G4}$ from -0.9~V to -0.7~V, while at the same time making $V_{G2}$ more negative. 
The net effect is to strengthen the electrostatic confinement along the sample $x$ direction. 
For the case of $V_{G4}=-0.7$~V, the maximum principal $g$-factor ($g_3$) is 1.7$\pm$0.1, while the minimum principal $g$-factor ($g_1$) is 0.3$\pm$0.2.  
By comparing the 3D $g$-tensor surfaces in Figure~\ref{fig:Figure2}a-ii and Figure~\ref{fig:Figure2}b-ii it is clear that both the size and orientation of the $g$-tensor are sensitive to $V_{G4}$.
To demonstrate the observed change in $g$-tensor orientation, Figure~\ref{fig:Figure2}b-iii reproduces the $g$-factor around the \RR{$y$} axis.
For $V_{G4}=-0.7$~V we highlight that the tilting into the sample $xz$ plane is now $\phi_0=70^{\circ}\pm3^{\circ}$, compared to $\phi_0=42^{\circ}\pm2^{\circ}$ $V_{G4}=-0.9$~V.  
The observation that the orientation of the principal axes is strongly affected by the gate bias, even for a single hole, is the key result of the $g$-tensor data set presented in Figure \ref{fig:Figure2}b.

\begin{center}
	\begin{tabular}{|c|c|c|} 
		\hline
		\hline
		Parameter	& $V_{G4}=-0.9$~V  & $V_{G4}=-0.7$~V \\
		\hline
		$g_1$	& 1.4$\pm$0.2	& 0.3$\pm$0.2	\\
		\hline
		$g_2$	& 2.3$\pm$0.1	& 1.0$\pm$0.1	\\
		\hline
		$g_3$	& 3.9$\pm$0.1	& 1.7$\pm$0.1	\\
		\hline
		\hline
		$\vartheta_0$	& 19$^\circ\pm 7^{\circ}$	& $10^\circ\pm 9^{\circ}$	\\
		\hline
		$\phi_0$	& $42^\circ\pm 2^{\circ}$	& $70^\circ\pm 3^{\circ}$	\\
		\hline
		$\theta_0$	& $9^\circ\pm 3^{\circ}$	& $-20^\circ\pm 6^{\circ}$	\\
		\hline
		\hline
	\end{tabular}
	\captionof{table}{Principle $g$-factors for a single hole in a silicon quantum dot, measured at two different confinement ($V_{G4}$) potentials. The values are extracted by fitting the respective data set in Figure~\ref{fig:Figure2} to Equation (1). See the Appendix E for the fitting procedure.  
	}
	\label{table:FittingParameters}
\end{center}

\subsection{Numerical simulations of the single-hole $g$-tensor}

\begin{figure*}[!bt]
	\centering
	\includegraphics[width=0.9\columnwidth]{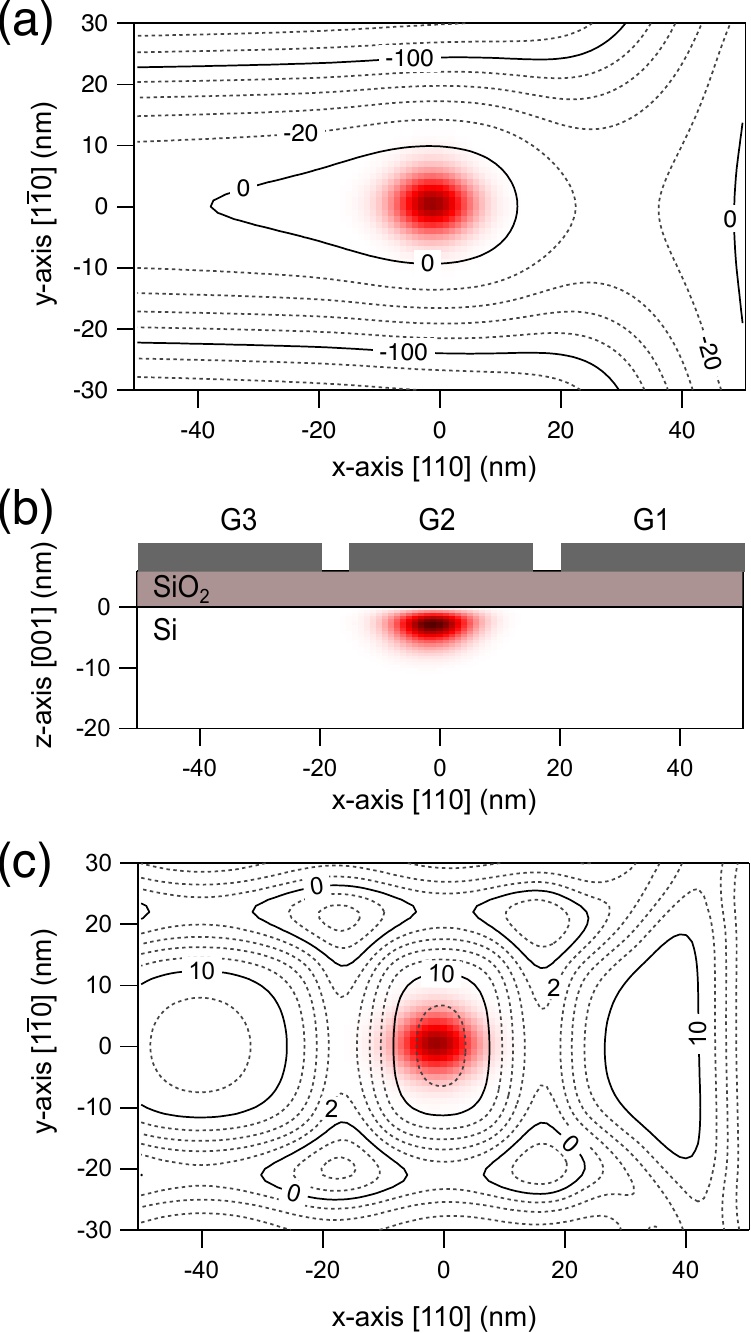}
	\caption[]{\small Numerical simulations of the electrostatic confinement, strain, and single hole eigenstates. 
	(a) Electrostatic confinement profile (contour lines) and hole ground state probability density (colour map) in the $xy$ plane. 
	The contours are spaced by 20~meV. 
	The in-plane confinement is strongest along the sample $y$-axis [1$\overline{1}$0], consistent with the strong influence of the large C-gate. 
	Confinement along the sample $x$-axis (crystal [110] axis) is weaker, and is provided by gates G1 and G3. 
	This simulation is for the experimentally applied voltages, as described in the methods.  
	(b) Hole probability density in the sample $xz$ plane. 
	The footprint of each gate is shown (the vertical height of the Al gates is $30+$~nm).
	(c) Spatial profile of the strain-induced HH-LH splitting in energy, with the hole probability density overlaid as a red colour map and HH-LH splitting shown as contours spaced by 2~meV. 
	}
	\label{fig:Figure3}
\end{figure*} 

To explore the physical origins of the $g$-tensor magnitude and orientation, we perform detailed three-dimensional modelling of the device, including (i) the real lithographic gate stack determined from design and device microscopy, (ii) strain built-up accompanying cool-down to cryogenic temperatures, (iii) self-consistent electrostatics with holes accumulating at the Si/SiO$_2$ interface at experimentally applied gate biases, and (iv) quantum mechanics of the Si complex valence band (parameterized via the $6\times6$ Luttinger-Kohn Hamiltonian with Bir-Pikus coupling to the lattice strain).  

In systems with strong spin-orbit coupling, the confinement and spin properties are inextricably linked. 
Therefore we begin the modelling of the hole $g$-tensor by first investigating the hole confinement profile.
Figure~\ref{fig:Figure3}a shows the calculated electrostatic confinement potential in the $xy$ plane. 
The hole ground state probability density projected to the same plane is overlaid as a colour map in Figure~\ref{fig:Figure3}a. 
Similarly, Figure~\ref{fig:Figure3}b shows the hole probability density projected to the sample $xz$ plane.
The hole ground state is mostly Heavy-Hole (HH) in character. 
The vertical extent of the wave function is $\sim7$~nm, and the diameter is $\sim30$~nm, consistent with the diameter estimated from the measured charging energy in Figure \ref{fig:StabilityDiagramFigure}d (see S6 or Ref. \citenum{supp}).
Therefore the holes are confined in a thin disk-like wavefunction, which is pulled tightly against the Si/SiO$_2$ interface. 
The axis of strongest orbital confinement is out-of-plane with respect to the Si/SiO$_2$ interface.

Holes confined to a simple 2D-like geometry will have the primary magnetic axis aligned with the axis of strongest confinement\cite{winkler2003spin,wang2016anisotropic,
bogan2017consequences}. 
One might therefore expect our disk-like hole wavefunction to have the largest $g$-factor for an out-of-plane magnetic field, corresponding to $\phi_0~\approx~0$.  
However, the experimental results in Figure~\ref{fig:Figure2} show that the largest $g$-factor is strongly tilted away from the axis of strongest confinement, with $\phi_0$~>40$^\circ$.

A non-zero $\phi_0$ could, in principal, be caused by a drastic rotation of the axis of strongest confinement, due to a complete change in the electrostatic confinement potential (see S3 or Ref. \citenum{supp}).
However detailed numeric simulations show that no reasonable range of gate voltages, interface steps, or surfaces charges can produce a substantial tilting of the out-of-plane confinement orientation. For all reasonable configurations, the single hole confinement in this MOS device is most strongly defined by the vertical hard wall potential of the Si/SiO$_2$ interface. 
Something other than electrostatic confinement is therefore needed to explain the non-zero $\phi_0$, and for this we turn to the impact of electrode-induced strain.

Strain develops in silicon MOS devices cooled to cryogenic temperatures \cite{thorbeck2015formation,park2016electrode} due to differences in the thermal contraction between metal electrodes and the silicon substrate. 
In particular, uniaxial strain alters the valence band Heavy-Hole-Light-Hole (HH-LH) splitting $\Delta_{\text{HH-LH}}$, while shear strains directly mix HH and LH components.
Both can have an enormous influence on the composition of the confined hole state and its spin properties\cite{venitucci2018electrical}. 
Figure~\ref{fig:Figure3}c shows the spatial profile of $\Delta_{\text{HH-LH}}$ in the active region of the device. 
The HH-LH splitting varies by over 50\% across the device, and follows the lithography of the aluminum gate stack. 
The strain varies most rapidly at the edges and corners of the metal gates; shear strains concentrate there as well. 
Under the gates, biaxial compression by the shrinking metal pushes the LH basis states deeper in energy relative to HH states, i.e., acting in the same direction as the out-of-plane confinement. 
The impact of electrode-induced strain is particularly strong in these silicon MOS devices since the electrodes are separated by only 5.9~nm from the active charge region.

\begin{figure*}[!hbt]
	\centering
	\includegraphics[width=2\columnwidth]{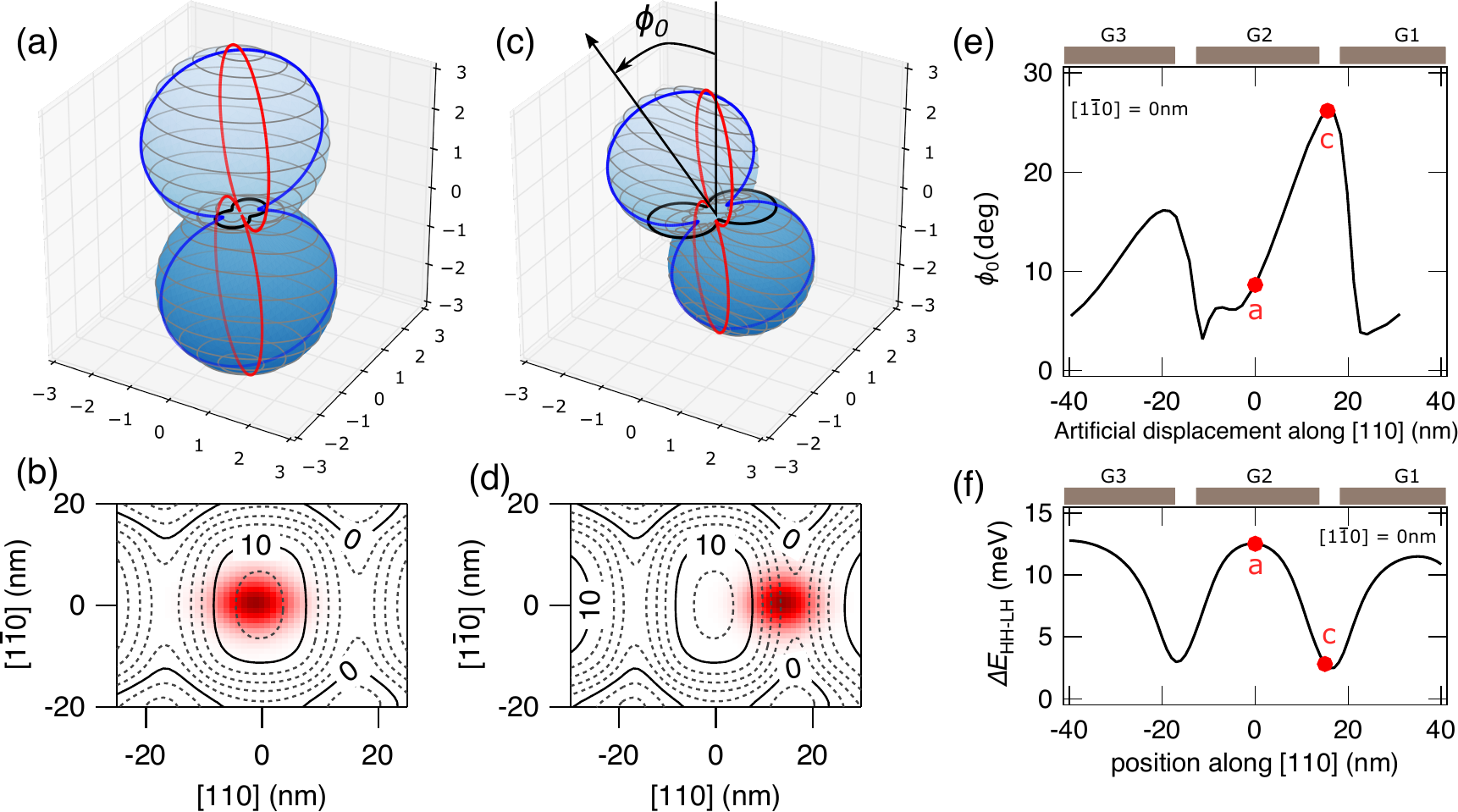}
	\caption[]{\small Tilted $g$-tensor and electrode-induced strain. 
		(a) Simulated $g$-tensor surface for the wavefunction position and strain profile indicated in panel (b). 
		The hole is localised directly below the G2 gate, where the local strain gradients are low and shear strains are minimal. 
		The orientation of the $g$-tensor is primarily defined by the orientation of the confinement, with the largest $g$-factor occurring for approximately out-of-plane magnetic field, more details in (e).
		The intersection of the $g$-tensor surface with the sample $xy$ (black), $xz$ (blue) and $yz$ (red) planes is indicated by the respective solid lines. 
		(c) Same as (a), except forcing hole to the position indicated in (d).
		The contour lines in (b) and (d) are spaced by 2~meV.
		(e) Shows the $g$-tensor tilting ($\phi_0$) as a function of the artificial shift of the electrostatic confinement along the sample $x$-axis.
		Red circles indicate the $x$-axis position and $\phi_0$ value for the case in (a) and (c).
		Above the figure we schematically indicate the location of gates G1, G2, and G3. 
		We find that the regions near gate edges correspond to the regions of largest tilting $\phi_0$. 
		(f) Shows the simulated strain-induced $\Delta_{\text{HH-LH}}$ for a line cut along the $x$-axis. 
		The $\Delta_{\text{HH-LH}}$ is not symmetric along sample $x$-axis, since the lithography \RR{and confinement potential} are not symmetric along the sample $x$-axis \RR{(see Figure \ref{fig:Figure3}).}
	}
	\label{fig:Figure4}
\end{figure*}

In an ideal device, the hole lies directly below the centre of the G2 gate, as shown in Figure~\ref{fig:Figure3}. 
Figure~\ref{fig:Figure4}a presents the simulated $g$-tensor surface of the single hole in its ground state (shaded blue surface). 
In this configuration the $g$-tensor is as expected for the predominantly HH-like state --- the largest $g$-factor occurs for a nearly out-of-plane magnetic field, with small but nonzero transverse components and tilt $\phi_0$ of $8^{\circ}$ due to the non-zero LH admixture. 
To tilt the hole $g$-tensor significantly out of the 2D plane (i.e. $ \phi_0 >10^{\circ}$) it is necessary to displace the hole wavefunction away from the point of near symmetry that occurs directly under a gate. This displacement can be due to atomic steps, surface charges, or other fluctuations of the Si/SiO$_2$ interface. 
A wavefunction displacement is also realizable experimentally by altering the different gate biases, such as $V_{G4}$. 
Figure \ref{fig:Figure4}c presents the $g$-tensor surface simulated when the hole wavefunction is electrostatically displaced by about 15~nm to a region of highly non-uniform strain, as indicated in Figure \ref{fig:Figure4}d. 
The $g$-tensor surface in Figure \ref{fig:Figure4}c is clearly tilted away from the Si/SiO$_2$ interface with $\phi_0>26^\circ$. 

Figure \ref{fig:Figure4}e shows the extracted tilt angle $\phi_0$ of the $g$-tensor as the hole is artificially being forced to various points along the sample $x$-axis. 
The spatial dependence on $\phi_0$ tracks the strain profile, responsible also for $\Delta_{\text{HH-LH}}$, shown in Figure~\ref{fig:Figure4}f. 
When Bir-Pikus strain terms were omitted from the numerical model the $g$-tensor tilting is suppressed, with a maximum $\phi_0<1^{\circ}$. 
These simulations suggest that it is the non-uniform strain profile that causes the observed orientation of the $g$-tensor to be misaligned from the electrostatic confinement orientation. In addition, \RR{residual strains associated with cryogenic cooling and/or processing} have been discussed as a likely mechanism causing a discrepancy between the calculated and observed g-tensor of holes in SOI nanowire quantum dots\cite{venitucci2018electrical}.

Other plausible mechanisms, such as HH-LH mixing by the microscopically low-symmetry Si/SiO$_2$ interface\cite{IKR1996}, were evaluated and deemed insufficiently strong to significantly renormalize the hole state $g$-tensor (see S2 of Ref. \citenum{supp}).

\section{CONCLUSIONS}
In this work we have experimentally studied the 3D $g$-tensor of a single hole in a silicon MOS based quantum dot. 
We characterised the full 3D $g$-tensor for two different bias configurations. 
Our results demonstrate strong electric control over both the magnitude and orientation of the single hole $g$-tensor. 

A key experimental result is the wide range of control over the $g$-factor, particularly the ability to configure a "sweet spot" where most components of $dg/dV$ approach zero. For spin qubits, the coupling between hole spins and electric fields is a balancing act, where some large component of $dg/dV$ maximises the EDSR Rabi frequency. 
However, a large $dg/dV$ also amplifies the impact of electrical noise leading to a shorter $T^*_2$. In the device under study we show that over a small range of bias configurations the hole can be tuned in-situ to a region of high $dg/dV$, which is ideal for rapid spin manipulation, then to a sweet spot where $dg/dV=0$ for a dominant gate, which is ideal for long lifetime qubit storage, prior to qubit readout\cite{froningultrafast}.

Since these results are for a single hole, in the lowest quantum dot orbital state, it is possible to compare the experimental data with detailed theoretical models. 
These models suggest that electrode-induced strain has a significant effect on hole-spin states in p-SiMOS based quantum dots, and show how the effects of strain vary dramatically as the hole wavefunction is moved around with gate biases.

We conclude that the effects of \textit{non-uniform} strain are critical for understanding the single hole g-tensor, particularly in MOS devices. 
Furthermore, the impact of \textit{non-uniform} electrode induced strain is relevant to a wide variety of hole-spin based devices.
Therefore, these results raise an interesting question: How effectively can the spatial strain profile be engineered to optimise the performance of hole spin based devices?

Finally, non uniform electrode-induced strain has not previously been considered as a mechanism enabling all electric spin manipulation of hole-based qubits. 
While overall strain has been used in spin-qubit devices\cite{dobbie2012ultra,hendrickx2018gate}, it has typically been used to engineer the static isotropic spin properties, particularly the HH-LH splitting. 
This work points to a potentially new technology for spin-qubits, where specific gate geometries are designed to engineer \emph{non-uniform} strain for optimised speed and performance. 


\section*{ACKNOWLEDGMENTS}
This work was funded by the Australian Research Council (DP150100237, DP200100147, and FL190100167) and the US Army Research Office (W911NF-17-1-0198). Devices were made at the NSW node of the Australian National Fabrication Facility. D.S.M. acknowledges the support by the Georg H. Endress foundation. K.M.I. acknowledges support from a Grant-in-Aid for Scientific Research by MEXT.  T.D.L. acknowledges support from the Gordon Godfrey Bequest Sabbatical grant.

\section*{APPENDIX}

\subsection*{Appendix A: Sample Details}
The device studied in this work was fabricated using the same processing procedure, but in a different processing run, as previous planar silicon hole quantum dots devices\cite{liles2018spin}. 
During operation the R-gate is negatively biased to accumulate a 2D hole gas at the Si/SiO$_2$ interface below. 
A single quantum dot is defined by positively biasing gates G1, G3, G4, and the C-gate. 
G2 acts as the dot plunger gate and is operated in the negatively biased regime. 
If G4 is made sufficiently negative the device forms a double dot (see S7 of Ref. \citenum{supp}). 
Full voltages are provided in S1 of Ref. \citenum{supp}. 
The charge sensor is operated by negatively biasing the sensor top gate to facilitate hole transport.
Sensor barrier gates, which are fabricated underneath the sensor top gate, are positively biased to form a region of high $dI_{\text{sens}}/dV$ (either a quantum dot or sharp pinch-off) which is used to charge sense the quantum dot below G2. 
We confirmed that the device operates down to a single hole using tunnel rate independent measurements. 
\\
\subsection*{Appendix B: Measurement Details}
All measurements were performed in a BlueFors XLD dilution refrigerator with a base temperature of 20~mK. 
For charge sensor measurements we monitor $dI_{\text{sens}}/dV_{G2}$ using the standard dual lock-in technique with dynamic feedback to optimise the charge sensor signal\cite{yang2011dynamically}. 
The charging energy and the $g$-factors presented in Figure~\ref{fig:StabilityDiagramFigure} and Figure~\ref{fig:Figure2} were extracted from the spacing in $V_{G2}$ between the $N=1$ and $N=2$ Coulomb peaks\cite{kouwenhoven2001few}. The spacing in $V_{G2}$ was then converted into energy using the lever arm, $\Delta E = \alpha_{G2} \Delta V_{G2} $, where $\alpha_{G2}$ = (0.166$\pm$0.007) eV/V. 
The full data set is presented in S4 of Ref. \citenum{supp}. 
We have confirmed that the lever arm is independent of the gate voltages within the operating range of the experiment. 
For all rotation measurements we first confirm that $\Delta E (B)$ is linear in B up to 1~T for all directions.
\\
For the rotation matrices in Equation (\ref{eqn:gExp}) we use the definition
\begin{equation}
     R(\vartheta_0,\phi_0,\theta_0) =  R_z(\theta_0) R_y(-\phi_0) R_z(\vartheta_0)
     \tag{B.1}
\end{equation}
where $R_y$ and $R_z$ are the standard 3D rotation matrices around $y$ and $z$ axes.
\\
\subsection*{Appendix C: Multiscale device modelling including strain} 
For modeling the hole states, we use a custom numerical framework for the construction and multiscale simulations of the three-dimensional multilayer and multimaterial device model. 
Layout construction from production masks, with attention to their orientation with respect to the principal axes of the silicon substrate, is augmented by TEM data, process models, and known details of fabrication steps. 
To account for the stress which builds up when cooling the heterogeneous system to cryogenic temperatures, we solve the stationary stress-strain problem for the entire layout,  assuming the device is unstrained at the end of fabrication and that the device is free of any cracks or voids.  

The obtained strain pattern can be combined with self-consistent Schr\"odinger-Poisson calculations, using a Thomas–Fermi approximation to model the partitioned two-dimensional hole gas accumulated outside the quantum region.  
The gate potentials for Schr\"odinger-Poisson are taken from those employed in the operation regime, including a global offset. 
Next, the single quantized hole in a Si complex valence band is treated quantum-mechanically by solving, in 3 dimensions, the 6$\times$6 Luttinger-Kohn Hamiltonian with Bir-Pikus strain terms, subject to realistic electrostatic confinement and strain. 
We extract the $g$-tensor by evaluating splittings of the ground (or excited) state doublet by the magnetic field at various angles relative to the simulated device.

In addition, modelling of the single hole state has been conducted for a hypothetical 3D harmonic confinement. The hypothetical modelling is illuminating due to its much simplified parameter space of only three confinement strengths and three rotation angles. Further details are presented in S2 and S3 of Ref. \citenum{supp}.  
\\
\subsection*{Appendix D: Estimation of minimum Rabi frequency} To estimate a minimum Rabi frequency we consider a 200mT magnetic field, and a 4mV AC signal applied to the gate G4. We use Equation 3 from Ref. \citenum{crippa2018electrical}
We simplify this estimation by considering only the terms ${g_y}$ and d${g_y}$/d$V_{G4}$ (ie naively considering that the tensor $\tensor{g}$ and $d\tensor{g}/dV_{G4}$ are diagonal in the measurement frame).
\\
\end{multicols}

\noindent\rule[0.5ex]{\linewidth}{0.5pt}
\subsection*{Appendix E: $g$-tensor definition and fitting procedure}
In this work we define the 3$\times$3 $g$-tensor, $\tensor{g}$, using six parameters parameters ($g_1, g_2, g_3, \vartheta_0, \phi_0$, $\theta_0$) such that
\begin{equation}
\tensor{g} = R_{z}(\vartheta_0)R_{y}(-\phi) R_{z}(\theta_0) 
\begin{pmatrix}
g_{1}& 0 & 0 \\
0 & g_{2} & 0 \\
0 & 0 & g_{3}
\end{pmatrix} R_{z}^{-1}(\theta_0) R_{y}^{-1}(-\phi_0) R_{z}^{-1}(\vartheta_0) 
\tag{E.1}
\label{eqn:gExpAPP}
\end{equation}
where $g_{1}$, $g_{2}$ and $g_{3}$ are the principle $g$-factor values, $\theta_0$, $\phi_0$, and $\vartheta_0$ define free rotations of the matrix allowing arbitrary orientation of the principle magnetic axes with respect to the sample ($x$,$y$,$z$) axes (the external frame of reference). Here $R_{y}$ and $R_{z}$ are the standard rotation matrices around the $y$ and $z$ planes respectively. Our convention is to use $R_{y}(-\phi_0)$ for rotation around the $y$ axis.

To extract the hole $g$-tensor the full experimental data set is simultaneously fit to Equation \ref{eqn:gExpAPP}. Below is the procedure used to fit all data to Equation \ref{eqn:gExpAPP}. The procedure input takes experimental data points in a .csv file with three columns; (1) $\theta$, (2) $\phi$, (3) observed $g$-factor, where $\theta$ and $\phi$ define the applied magnetic field orientation (see main text Figure 2). Prior to fitting we apply the condition that $g_1 < g_2 < g_3$. The best fit values of the $g$-tensor free parameters ($g_1, g_2, g_3, \vartheta_0, \phi_0, \theta_0$) are presented in Table 1 of the main text.\\
\noindent\rule[0.5ex]{\linewidth}{0.5pt}
\begin{multicols}{2}
\subsection*{Appendix F: Code for g-tensor fitting}
The python notebook used for fitting the experimental g-factor data to Equation \ref{eqn:gExpAPP} can be found on the GitHub Repository here: https://github.com/ScottDLiles/gFactorFitting.git


\small

\end{multicols}

\end{document}